\documentclass[twocolumn,aps,showpacs,preprintnumbers,amsmath,amssymb, superscriptaddress]{revtex4-1}
\pdfoutput=1

\usepackage{bm}
\usepackage{graphicx}
\usepackage{color}
\usepackage{xspace}

\renewcommand{\vec}{\mathbf}

\begin{document}

\title{Observation of strong-coupling pairing with weakened Fermi-surface nesting at optimal hole doping in Ca$_{0.33}$Na$_{0.67}$Fe$_2$As$_2$}

\author{Y.-B. Shi}
\affiliation{Beijing National Laboratory for Condensed Matter Physics, and Institute of Physics, Chinese Academy of Sciences, Beijing 100190, China}
\author{Y.-B. Huang}
\affiliation{Beijing National Laboratory for Condensed Matter Physics, and Institute of Physics, Chinese Academy of Sciences, Beijing 100190, China}
\affiliation{Paul Scherrer Institut, Swiss Light Source, CH-5232 Villigen PSI, Switzerland}
\author{X.-P. Wang}
\affiliation{Beijing National Laboratory for Condensed Matter Physics, and Institute of Physics, Chinese Academy of Sciences, Beijing 100190, China}
\affiliation{Paul Scherrer Institut, Swiss Light Source, CH-5232 Villigen PSI, Switzerland}
\author{X. Shi}
\affiliation{Beijing National Laboratory for Condensed Matter Physics, and Institute of Physics, Chinese Academy of Sciences, Beijing 100190, China}
\affiliation{Paul Scherrer Institut, Swiss Light Source, CH-5232 Villigen PSI, Switzerland}
\author{A. van Roekeghem}
\affiliation{Beijing National Laboratory for Condensed Matter Physics, and Institute of Physics, Chinese Academy of Sciences, Beijing 100190, China}
\affiliation{Centre de Physique Th\'{e}Žorique, Ecole Polytechnique, CNRS-UMR7644, 91128 Palaiseau, France}
\author{W.-L. Zhang}
\affiliation{Beijing National Laboratory for Condensed Matter Physics, and Institute of Physics, Chinese Academy of Sciences, Beijing 100190, China}
\author{N. Xu}
\affiliation{Beijing National Laboratory for Condensed Matter Physics, and Institute of Physics, Chinese Academy of Sciences, Beijing 100190, China}
\affiliation{Paul Scherrer Institut, Swiss Light Source, CH-5232 Villigen PSI, Switzerland}
\author{P. Richard}\email{p.richard@iphy.ac.cn}
\affiliation{Beijing National Laboratory for Condensed Matter Physics, and Institute of Physics, Chinese Academy of Sciences, Beijing 100190, China}
\affiliation{Collaborative Innovation Center of Quantum Matter, Beijing, China}
\author{T. Qian}\email{tqian@iphy.ac.cn}
\affiliation{Beijing National Laboratory for Condensed Matter Physics, and Institute of Physics, Chinese Academy of Sciences, Beijing 100190, China}
\author{E. D. L. Rienks}
\affiliation{Helmholtz-Zentrum Berlin, BESSY, D-12489 Berlin, Germany}
\author{S. Thirupathaiah}
\affiliation{Helmholtz-Zentrum Berlin, BESSY, D-12489 Berlin, Germany}
\affiliation{IFW-Dresden, P.O.Box 270116, D-01171 Dresden, Germany}
\author{K. Zhao}
\affiliation{Beijing National Laboratory for Condensed Matter Physics, and Institute of Physics, Chinese Academy of Sciences, Beijing 100190, China}
\author{C.-Q. Jin}
\affiliation{Beijing National Laboratory for Condensed Matter Physics, and Institute of Physics, Chinese Academy of Sciences, Beijing 100190, China}
\affiliation{Collaborative Innovation Center of Quantum Matter, Beijing, China}
\author{M. Shi}
\affiliation{Paul Scherrer Institut, Swiss Light Source, CH-5232 Villigen PSI, Switzerland}
\author{H. Ding}
\affiliation{Beijing National Laboratory for Condensed Matter Physics, and Institute of Physics, Chinese Academy of Sciences, Beijing 100190, China}
\affiliation{Collaborative Innovation Center of Quantum Matter, Beijing, China}

\date{\today}

\begin{abstract}
We report an angle-resolved photoemission investigation of optimally-doped Ca$_{0.33}$Na$_{0.67}$Fe$_2$As$_2$. The Fermi surface topology of this compound is similar to that of the well-studied Ba$_{0.6}$K$_{0.4}$Fe$_2$As$_2$ material, except for larger hole pockets resulting from a higher hole concentration per Fe atoms. We find that the quasi-nesting conditions are weakened in this compound as compared to Ba$_{0.6}$K$_{0.4}$Fe$_2$As$_2$. As with Ba$_{0.6}$K$_{0.4}$Fe$_2$As$_2$ though, we observe nearly isotropic superconducting gaps with Fermi surface-dependent magnitudes. A small variation in the gap size along the momentum direction perpendicular to the surface is found for one of the Fermi surfaces. Our superconducting gap results on all Fermi surface sheets fit simultaneously very well to a global gap function derived from a strong coupling approach, which contains only 2 global parameters. 
\end{abstract}



\maketitle

Although there is a broad consensus on the existence of non-conventional superconductivity in the Fe-based superconductors, intense debates persist on the precise nature of the pairing mechanism in these compounds. Arguably, the candidate models can be divided into two main categories. On one side, the Fermi surface (FS)-driven pairing mechanisms or weak-coupling approaches consider as primordial the interactions between the various FSs \cite{Graser_NJP2009,Hirschfeld_RoPP2011,KontaniPRL2010}. On the other side, short-range pairing mechanisms, which may include approaches from the intermediate to the strong coupling \cite{SeoPRL2008,C_Fang_PRX2011,Y_Zhang_PRL2010,HuJP_PRX2012,HuJP_SR2012}, are more naturally described in the real space. These various models necessarily have fingerprints in the size and symmetry of the superconducting (SC) order parameter, which are accessible directly in the momentum space by angle-resolved photoemission spectroscopy (ARPES).

The easiest way to address this debate using a momentum-resolved probe such as ARPES is to tune the size of the various FSs, which is done by varying the carrier concentration in the Fe-As planes. The 122 family of ferropnictides is ideal for this purpose since it counts several members with relatively high SC critical temperatures ($T_c$'s), in addition to leave nice and shinny cleaved surfaces. In this paper, we focus on optimally-doped Ca$_{0.33}$Na$_{0.67}$Fe$_2$As$_2$, which has a $T_c$ nearly as high as the well-studied Ba$_{0.6}$K$_{0.4}$Fe$_2$As$_2$ compound, despite a much larger hole concentration per Fe at optimum concentration \cite{K_Zhao_PRB84}. Consistently, we observe very large hole FSs centered at the Brillouin zone (BZ) center ($\Gamma$ point) showing weaker quasi-nesting conditions to the M-centered electron FSs than in Ba$_{0.6}$K$_{0.4}$Fe$_2$As$_2$ \cite{Ding_EPL,L_ZhaoCPL25}. We determine a rather isotropic SC gap that is FS-dependent, with SC gap sizes ranging from 5.7 to 9.7 meV. Interestingly, we show that the SC gap on one of these bands is slightly modulated as a function of the out-of-plane momentum $k_z$. More importantly, we demonstrate that the SC gap data on all the FSs can be fit all together using the same global gap function derived from a strong coupling, with the same 2 global gap parameters, thus suggesting that local interactions play a major role in the Cooper pairing in the Fe-based superconductors. 

Single-crystals of Ca$_{0.33}$Na$_{0.67}$Fe$_2$As$_2$ showing bulk superconductivity at $T_c = 33$ K were grown by the self-flux method \cite{K_Zhao_JPCM2}. Most of the synchrotron ARPES measurements were performed at beamline SIS of the Swiss Light Source. Synchrotron data were also recorded at 1 K using the 1-cubed ARPES end-station of BESSY, and additional SC gap measurements were performed in our own facilities at the Institute of Physics, Chinese Academy of Sciences, using the He $\alpha$I line of an helium discharge lamp. All these systems are equipped with VG-Scienta R4000 electron analyzers and the angular resolution was set to 0.2$^{\circ}$ while the energy resolution for the SC gap measurements ranged from 4 to 7 meV. All samples were cleaved \emph{in situ} and measured in a working vacuum better than 5$\times$10$^{-11}$ torr. In the following, we label the in-plane momentum values with respect to the 1 Fe/unit cell Brillouin zone (BZ) and use $c'=c/2$ as the distance between two Fe planes. 

\begin{figure}[!t]
\begin{center}
\includegraphics[width=8.5cm]{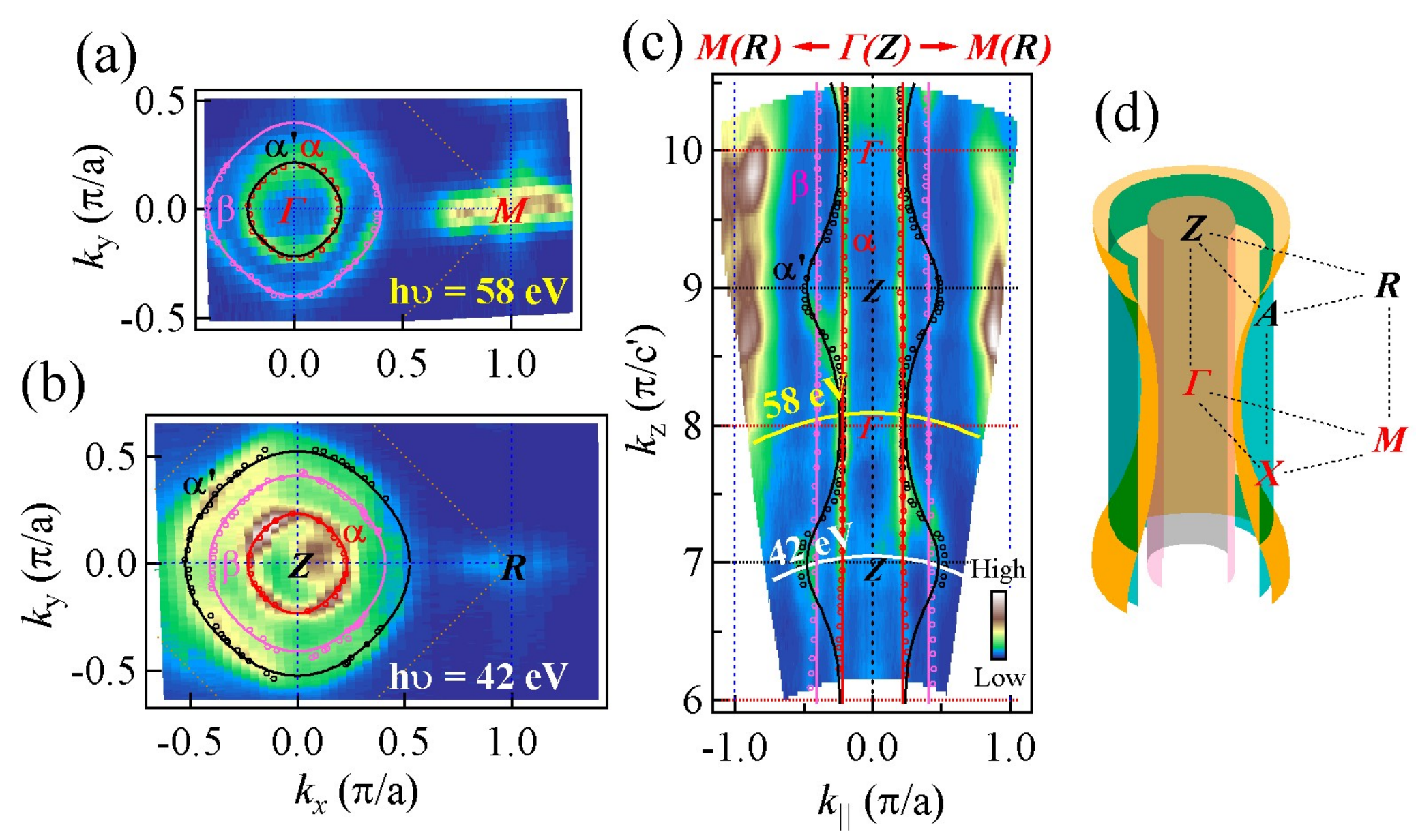}
\end{center}
\caption{\label{fig1_FS}(Color online) (a) and (b) ARPES intensity plots ($\pm 5$ meV integration) of the FS of Ca$_{0.33}$Na$_{0.67}$Fe$_2$As$_2$ recorded with 58 eV ($k_z\approx 0$) and 42 eV ($k_z\approx \pi/c'$) photons, respectively. Open symbols represent the $k_F$ positions obtained from MDC fitting at $E_F$ for the $\alpha$, $\alpha$' and $\beta$ bands, and the solid lines correspond to fits of these $k_F$ positions to the function form $k_F^{\alpha, \alpha', \beta}=k_{F,0}^{\alpha, \alpha', \beta}+k_{F,\phi}^{\alpha, \alpha', \beta}\cos(4\phi)$. (c) ARPES intensity plot ($\pm 5$ meV integration) of the $k_z$ momentum dispersion at $E_F$ obtained by converting the photon energy into $k_z$ using the free-electron approximation \cite{DamascelliPScrypta2004} with an inner potential of $V_0=12.7$ eV. The $k_F$ positions are indicated by open symbols and solid lines are used as guides to the eye for the various FSs. Strong $k_z$ dispersion is observed only for the $\alpha'$ band within the $[\pi/2,3\pi/2]$ range with the form $k_F^{\alpha'}=k_0+k_{\perp}\cos(k_z\pi)$. The yellow and blue curved lines indicate the $k_z$ values for the mappings obtained at 58 eV and 42 eV, respectively. (d) 3D representation of the $\Gamma$-centered FSs.}
\end{figure}

To discuss the FS topology of Ca$_{0.33}$Na$_{0.67}$Fe$_2$As$_2$, we present in Figs. \ref{fig1_FS}(a) and \ref{fig1_FS}(b) ARPES intensity mappings of this material recorded with 58 eV and 42 eV photons, which correspond approximately to the $k_z=0$ and $k_z=\pi/c'$ planes, respectively. The FS topology, quite similar to that of other 122-ferropnictides \cite{RichardRoPP2011}, is composed of $\Gamma$(Z)-centered hole FS pockets and M(R)-centered electron FS pockets. Using the momentum distribution curves (MDCs) at the Fermi level ($E_F$), we extracted the Fermi wave vectors ($k_F$) of the hole FSs, as illustrated in Figs. \ref{fig1_FS}(a) and \ref{fig1_FS}(b). While we can distinguish only two hole FSs for $k_z\approx 0$, we clearly observe three for $k_z\approx\pi/c'$. Based on previous reports on similar materials \cite{RichardRoPP2011}, we conclude that two of them, the $\alpha$ and $\alpha'$ FSs, are nearly degenerate near $k_z=0$. 

The $\alpha'$ band, which is formed mainly by the even combination of the $d_{xz}$ and $d_{yz}$ orbitals \cite{XP_WangPRB85}, is the outermost FS near $k_z=\pi/c'$. As illustrated in Fig. \ref{fig1_FS}(c), this FS is the only one showing significant modulation along the $k_z$ axis, which we access by converting the probing photon energy into $k_z$ using the free-electron approximation \cite{DamascelliPScrypta2004} and an inner potential $V_0=12.7$ eV. This modulation is particularly obvious within the $[\pi/2,3\pi/2]$ range. As with optimally-doped Ba$_{0.6}$K$_{0.4}$Fe$_2$As$_2$ \cite{YM_Xu_NPhys2011}, the other bands do not exhibit any noticeable modulation with respect to $k_z$, and thus the three-dimensional (3D) FS of Ca$_{0.33}$Na$_{0.67}$Fe$_2$As$_2$ near the zone center, reproduced schematically in Fig. \ref{fig1_FS}(d), is essentially composed of cylinders, except for one strongly dispersive FS sheet.

\begin{figure}[!t]
\begin{center}
\includegraphics[width=8.5cm]{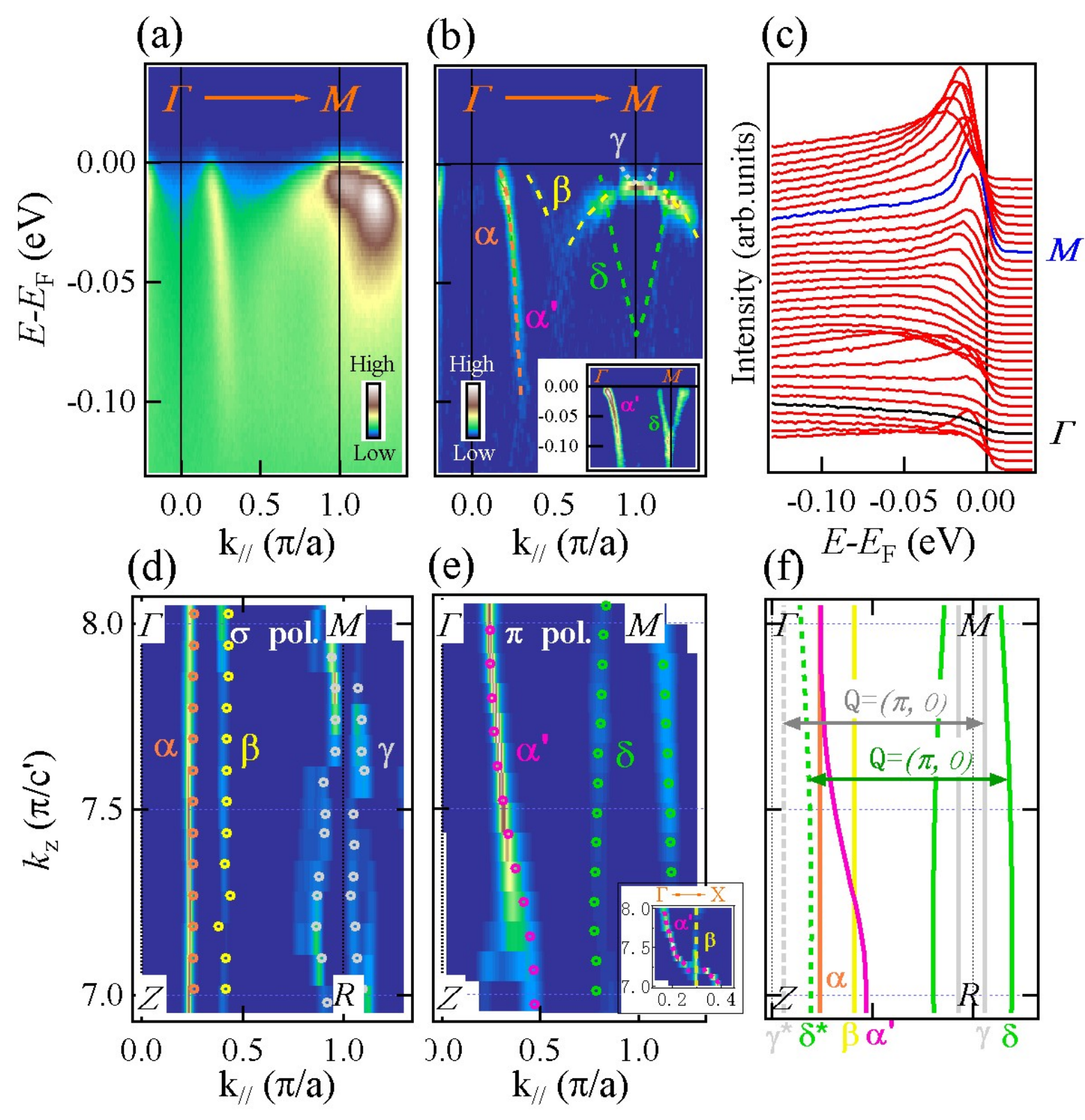}
\end{center}
\caption{\label{fig2_dispersion}(Color online) (a) ARPES intensity plot recorded at 10 K along $\Gamma$-M ($h\nu=63$ eV). (b) and (c) Corresponding intensity plot of 2D curvature \cite{P_Zhang_RSI2011} and EDC plot, respectively. Dashed lines in (b) are guides to the eye. The inset of (b) refers to data obtained with $\pi$ polarization, which show clearly the $\delta$ band. The EDCs at the $\Gamma$ and M points are respectively represented in black and blue in panel (c). (d) and (e) Intensity plots of 2D curvature in the $k_x-k_{z}$ plane recorded with $\sigma$ and $\pi$ polarizations, respectively. The inset in panel (e) is the 2D curvature intensity plot along the $\Gamma$-X direction recorded with $\pi$ polarized light. (f) Illustration of the nesting conditions along $\Gamma$(Z)-M(R). Solid lines represent the various FSs while the dashed lines correspond to FSs shifted by the antiferromagnetic vector $\vec{Q}$.}
\end{figure}

Figure \ref{fig2_dispersion} shows the electronic band dispersion along the $\Gamma$-M high-symmetry line. In addition to the $\Gamma$-centered hole bands discussed so far, the ARPES intensity plot along $\Gamma$-M in Fig. \ref{fig2_dispersion}(a), the corresponding intensity plot of 2D curvature \cite{P_Zhang_RSI2011} in Fig. \ref{fig2_dispersion}(b) and the EDC plot in Fig. \ref{fig2_dispersion}(c) illustrate well the M-centered electron pockets. To discuss the nesting conditions between the M-centered electron pockets and the $\Gamma$-centered hole pockets, which are central to FS-driven pairing mechanisms, we display in Figs. \ref{fig2_dispersion}(d) and \ref{fig2_dispersion}(e) the intensity plots of 2D curvatures in the $k_x-k_z$ plane, for data recorded using $\sigma$ and $\pi$ polarizations, respectively. As we can see from the summary of the FSs observed, which is illustrated in Fig. \ref{fig2_dispersion}(f), and as reported previously \cite{Evtushinsky_PRB87}, no elecron-hole pair of FS pockets has a good nesting. Although quasi-nesting in the sense of Ref. \cite{RichardRoPP2011} remains possible, it is weaker than in the Ba$_{1-x}$K$_{x}$Fe$_2$As$_2$ cousin compounds, despite a similarly high $T_c$.  

\begin{figure}[!t]
\begin{center}
\includegraphics[width=8.5cm]{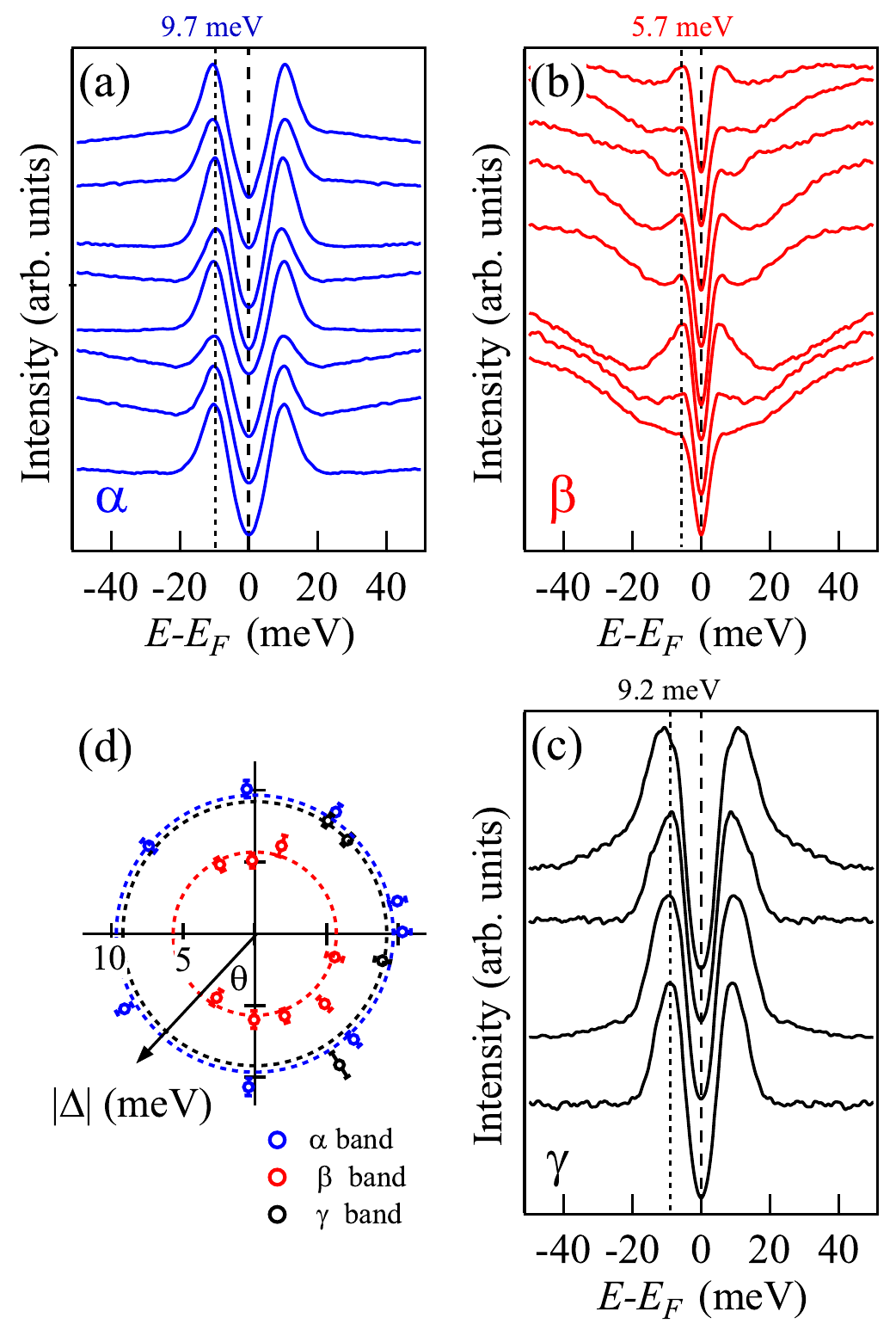}
\end{center}
\caption{\label{fig3_gap}(Color online) (a)-(c) Momentum distribution of the symmetrized EDCs, recorded at 12 K with the He $\alpha$I line (21.218 eV) of an helium discharge lamp, showing the SC gap along the $\alpha$, $\beta$ and $\gamma$ bands, respectively. The corresponding average gap values are $\Delta_{\alpha}=9.7$ meV, $\Delta_{\beta}=5.7$ meV and $\Delta_{\gamma}=9.2$ meV. (d) Polar distribution of the SC gap values obtained from panels (a)-(c).} 
\end{figure}

To check if the SC gap structure of Ca$_{0.33}$Na$_{0.67}$Fe$_2$As$_2$ is strongly affected by the relatively weaker quasi-nesting conditions as compared with Ba$_{0.6}$K$_{0.4}$Fe$_2$As$_2$, we performed ultra-high energy resolution measurements below the SC transition. The results are shown in Fig. \ref{fig3_gap}. Following a common practice, we symmetrized the EDCs in order to approximately remove the Fermi-Dirac cut-off. The momentum distribution of the resulting curves are given in Figs. \ref{fig3_gap}(a)-\ref{fig3_gap}(c) for the $\alpha$, $\beta$ and $\gamma$ FSs, respectively. As illustrated with the polar plot in Fig. \ref{fig3_gap}(d), rather isotropic SC gaps open on all these FSs, which is similar to observations on other 122-ferropnictides \cite{RichardRoPP2011} but contrasts with a recent report on Ca$_{1-x}$Na$_{x}$Fe$_2$As$_2$ where a noticeable gap anisotropy is found on the $\beta$ FS \cite{Evtushinsky_PRB87}. Interestingly, the gap found on the large $\beta$ FS is significantly smaller than on the $\alpha$ and $\gamma$ FSs. Similar observation has been reported from a calorimetric investigation of Ba$_{0.65}$Na$_{0.35}$Fe$_2$As$_2$ where $\Delta/k_BT_c$ ratios of 1.06 and 2.08 were found \cite{PramanikPRB84}. The gap sizes we determined directly from ARPES on Ca$_{0.33}$Na$_{0.67}$Fe$_2$As$_2$ are much larger though and more consistent with other Fe-based superconductors with comparable $T_c$'s. Indeed, while average SC gaps of $\Delta_{\alpha}=9.7$ meV ($\Delta_{\alpha}/k_BT_c=3.4$) and $\Delta_{\gamma}=9.2$ meV ($\Delta_{\gamma}/k_BT_c=3.2$) are found for the $\alpha$ and $\gamma$ FSs, the average SC gap size on the $\beta$ band is only $\Delta_{\beta}=5.7$ meV ($\Delta_{\beta}/k_BT_c=2$), a situation similar to that of Ba$_{0.6}$K$_{0.4}$Fe$_2$As$_2$, where the 6 meV SC gap found on the $\beta$ FS differs largely from the 12 meV SC gap found on the other bands \cite{Ding_EPL,L_ZhaoCPL25}. We note that our data show slightly larger SC gap sizes than in the study of Evtushinsky \emph{et al.} \cite{Evtushinsky_PRB87}, which is possibly due to a sample dependence.

\begin{figure}[!t]
\begin{center}
\includegraphics[width=8.5cm]{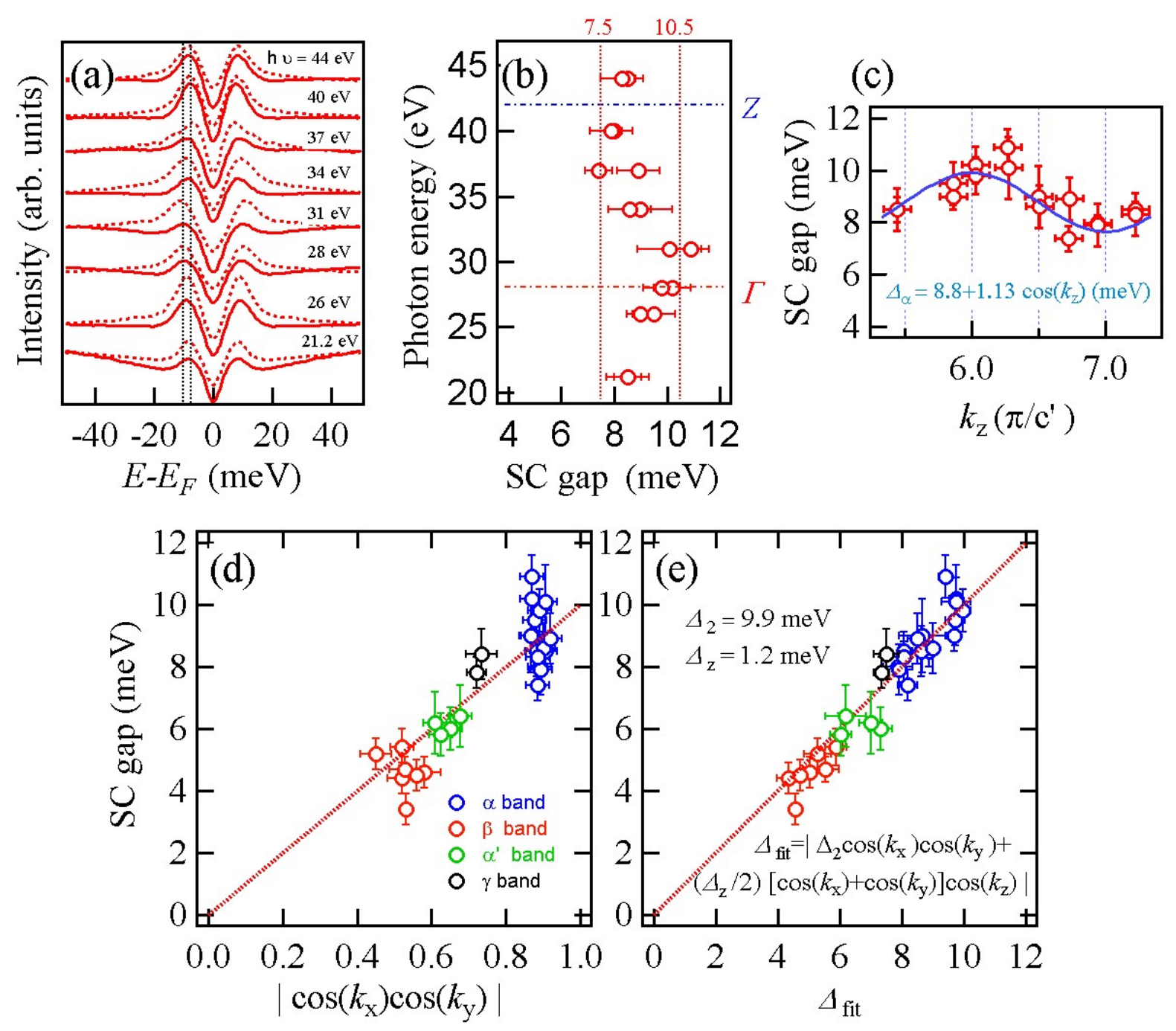}
\end{center}
\caption{\label{fig4_fit}(Color online) (a) Photon energy dependence of the symmetrized EDCs recorded at 1 K (except for the 21.2 eV data, recorded at 12 K) on the $\alpha$ FS. Solid and dashed curves correspond to two different $k_F$ points. The vertical dashed lines located at -10.5 meV and -7.5 meV are guides to the eye showing the dispersion of the peaks. (b) Photon energy dependence of the SC gap sizes extracted from (a). (c) Same as in (b) but with the photon energy converted into $k_z$. The data are fit to the function $\Delta_{\alpha}=8.8+1.13\cos(k_z)$ (meV). (d) SC gap magnitude on the various FSs as a function of the global gap function $|\cos(k_x)\cos(k_y)|$. (e) Same as (d) but for the gap function $\Delta_{\textrm{fit}}=|\Delta_2\cos(k_x)\cos(k_y)+(\Delta_z/2)[\cos(k_x)+\cos(k_y)]\cos(k_z)|$.} 
\end{figure}

A previous report on the cousin Ba$_{0.6}$K$_{0.4}$Fe$_2$As$_2$ optimally-doped compound showing non-negligible $k_z$ variations of the SC gap \cite{YM_Xu_NPhys2011}, especially on the $\alpha$ FS, we also recorded SC data at various photon energies for that particular FS. The corresponding symmetrized EDCs are displayed in Fig. \ref{fig4_fit}(a). The extracted gap size, shown in Fig. \ref{fig4_fit}(b), is slightly modulated with photon energy. More precisely, we found after converting the photon energy into $k_z$ that the gap size obeys the periodic function $\Delta_{\alpha}=8.8+1.13\cos(k_z)$ (meV), as shown in Fig. \ref{fig4_fit}(c). This result suggests the presence of interlayer interactions affecting the SC pairing. This effect is relatively small though as compared to the average value of the SC gap.

One of the main consequences of a short-range SC pairing mechanism is that the SC gap on \emph{each} of the various FS sheets obeys \emph{the same} global function defined in the entire momentum space. In other words, the SC gap magnitude depends essentially on the absolute $k_F$ position rather than on the relative positions of two FSs, in contrast to FS-driven pairing mechanisms for which inter-FS and intra-FS interactions dominate. Inelastic neutron scattering experiments suggest that the interactions between next-nearest Fe neighbors, characterized by the exchange parameter $J_2$, is the most important antiferromagnetic exchange parameter to describe the spin wave dispersion in the magnetically-ordered 122-ferropnictides. Assuming that fluctuations of these interactions persist in the SC materials and are responsible for the SC pairing leads to the conclusion that the SC pairing function must either have the form $\sin(k_x)\sin(k_y)$ or $\cos(k_x)\cos(k_y)$. We discard the former one since it implies the presence of nodes that are not detected in our experiments. In Fig. \ref{fig4_fit}(d), we plot the magnitude of the SC gap determined by ARPES as a function of the absolute value of the latter function, ARPES being directly sensitive only to the magnitude of the SC gap. The agreement is qualitatively pretty good with a linear relationship, with the data spreading over a considerable range of the [0,1] range limiting the $|\cos(k_x)\cos(k_y)|$ function.

The fit of the experimental data to a global gap function derived from short-range interactions using a single parameter for all the FS sheets pushes towards a local pairing origin in this particular material, similarly as in the other Fe-based superconductors \cite{Y_Huang_AIP2012}. The fit is not perfect though, suggesting that some parameters have been neglected. For instance, as mentioned above, there must exist a finite interlayer interaction responsible for the modulation of the SC gap on the $\alpha$ FS shown in Fig. \ref{fig4_fit}(c). Following previous studies on Ba$_{0.6}$K$_{0.4}$Fe$_2$As$_2$ \cite{YM_Xu_NPhys2011}, BaFe$_2$(As$_{0.7}$P$_{0.3}$)$_2$  \cite{Y_Zhang_NaturePhys2012} and Ba(Fe$_{0.75}$Ru$_{0.25}$)$_2$As$_2$ \cite{Nan_XuPRB87} addressing this issue for ARPES measurements of the SC gap on the Fe-based superconductors, we improve the gap function by including an interlayer coupling that translates into the global gap function $|\Delta_2\cos(k_x)\cos(k_y)+(\Delta_z/2)[\cos(k_x)+\cos(k_y)]\cos(k_z)|$, where we now have 2 parameters. The result of the fit, displayed in Fig. \ref{fig4_fit}(e), suggests the validity of this approach for Ca$_{0.33}$Na$_{0.67}$Fe$_2$As$_2$ as well, and thus suggests that this behavior is common to most of the 122 ferropnictide compounds. Interestingly, we extract the global parameters $\Delta_2=9.9$ meV and $\Delta_z=1.2$ meV, which leads to a $\Delta_z/\Delta_2$ ratio of 8.3 that is similar to the $J_2/J_z$ ratio of 7 determined from inelastic neutron scattering on the parent compound CaFe$_2$As$_2$ \cite{McQueeneyPRL2008}. Although interband and intraband interactions very near $E_F$ certainly play a very important role in the physics of the Fe-based superconductors, our current results add to previous ones suggesting that the SC pairing of electrons in these materials rather originates from short-range interactions better described in the real space.

We thank J.-P. Hu for useful discussions. This ARPES work was supported by grants from CAS (2010Y1JB6), MOST (2010CB923000, 2011CBA001000 and 2013CB921703), and NSFC (11004232, 11050110422, 11234014 and 11274362) of China, as well as by the Sino-Swiss Science and Technology Cooperation (project no. IZLCZ2 138954). This work was partly performed at the Swiss Light Source, Paul Scherrer Institut, Villigen, Switzerland, and at BESSY, Helmholtz Zentrum, Berlin, Germany.

\bibliography{biblio_long}

\end{document}